\documentclass[epj,twocolumn]{svjour}
\usepackage{graphics}
\usepackage[figuresright]{rotating}
\usepackage{amssymb}
\usepackage{bm}
\usepackage{epsfig}
\usepackage{color}
\usepackage{amsmath}
\usepackage{latexsym}

\newcommand{\beq}{\begin{eqnarray}}
\newcommand{\eeq}{\end{eqnarray}}

\begin{document}
\title{Growth Rate in the Dynamical Dark Energy Models}
\author{{Olga Avsajanishvili}\inst{1}, {Natalia A. Arkhipova}\inst{2}, {Lado Samushia}\inst{3,1}, and {Tina Kahniashvili}\inst{4,5,1}.
}                     
%
%
\institute{{Abastumani
Astrophysical Observatory, Ilia State University, 3-5 Cholokashvili
Ave., Tbilisi, 0194, Georgia}
\and {Astro Space Center of P.N.Lebedev Physical Institute, Russia, 84/32 Profsoyuznaya str.,
Moscow, 117997 Russia}
\and {Department of Physics, Kansas State University, 116 Cardwell Hall, Manhattan, KS, 66506, USA}
\and{McWilliams Center for Cosmology and Department of Physics, Carnegie Mellon University, 5000 Forbes Ave, Pittsburgh, PA 15213 USA}
\and {Department of Physics, Laurentian, University, Ramsey Lake Road, Sudbury, ON P3E 2C, Canada}}

\authorrunning{Avsajanishvili et al.}
\titlerunning{Growth rate in the Dynamical Dark Energy Models}

%
\date{\today}
%
\abstract{Dark Energy models with slowly-rolling cosmological scalar field
provide a popular alternative to the standard, time-independent
cosmological constant model. We study simultaneous evolution of
background expansion and growth in the scalar field model with the
Ratra-Peebles self-interaction potential. We use recent measurements
of the linear growth rate and the baryon acoustic oscillation peak
positions to constrain the model parameter $\alpha$ that describes
the steepness of the scalar field potential.
     } 


\maketitle

\section{Introduction}
\label{intro}
Cosmological observations now convincingly show that the expansion of the
Universe is now accelerating \cite{Weinberg:2012es}. One of the possible explanations of this
empirical fact is that the energy density of the Universe is dominated by so
called {\it dark energy} (DE) \cite{Yoo:2012ug}, a component with effective negative pressure.

The simplest DE candidate is a time-independent cosmological constant $\Lambda$, and the corresponding cosmological model, so called $\Lambda$CDM model, is considered to be a {\it concordance} model.
This simple model however suffers from fine turning and coincidence problems \cite{Martin:2012bt}.
In the attempt of constructing a more natural model of DE many alternative
scenarios have been proposed \cite{Caldwell:1997ii}.

One of the alternatives to a cosmological constant are the models of dynamical
scalar field. In these models a spatially uniform cosmological scalar field,
slowly rolling down it's almost flat self-interaction potential, plays a role of
time-dependent cosmological constant. This family of models avoids fine tuning
problem, having a more natural explanation for the observed
low energy scale of DE \cite{Samushia:2009dd,Zlatev:1998tr}. For the scalar field models (so called $\phi$CDM model)  the equation
of state $P_{\phi}= w\rho_{\phi}$ (with $P_\phi$ and $\rho_\phi$ pressure and energy density of the scalar field)
is time-dependent $w= w(t)$, and unlike the
cosmological constant $ w (t) \neq-1$, although at late-times it approaches -1.
When the scalar field energy density starts dominating the energy budget of the
Universe, the Universe expansion starts accelerating \cite{Bamba:2012cp,Bolotin:2011sz}.
Even though at low redshifts the predictions of
the model are very close to the ones of cosmological constant, the two models ($\Lambda$CDM and the dynamical DE model)
predict different observables over a wide-range of redshifts.

The scalar field models can be classified via their effective equation of state
parameter. The models with $ -1<w<-{1}/{3}$ are referred to as
quintessence models, while the models with $w<-1$ are referred as
phantom models. The quintessence models can be divided in two broad classes:
tracking quintessence, in which the evolution of the scalar field is slow, and
the thawing quintessence, in which the evolution is fast compared to the Hubble
expansion \cite{Caldwell:2005tm}.
In tracking models the scalar field exhibits tracking solutions in which
the energy density of the scalar filed scales as the dominant component at
the time, therefore the DE is subdominant but closely tracks first
the radiation and then matter for most of the cosmic evolution. At some point
in matter domination epoch the scalar field becomes dominant which results in its effective negative pressure and accelerated expansion \cite{Brax:2002vf}. The simplest
example of such a model is provided by a scalar field with an inverse-power-law
potential energy density $V_{\phi} \propto\phi^{-\alpha}$, $\alpha>0$ \cite{Ratra:1987rm}, so called {\it Ratra-Peebles} model.

The scalar field models predict a different background expansion history
and the growth rate compared to the cosmological constant model ones. Thus the scalar field model can be
distinguished from $\Lambda$CDM model through high precision measurements of distances and growth rates over a wide redshift range \cite{Samushia:2011cs,Pace:2011kb}.

In this paper we study generic predictions of slowly rolling scalar field
models by taking the Ratra-Peebles model as a representative example. We
present a self-consistent and effective way of solving the joint equations for
the background expansion and the growth rate. We use a compilation of recent
growth rate and baryon acoustic oscillation (BAO) peak measurements
to put constraints on the parameter $\alpha$
describing the steepness of the scalar field's potential.

This paper is organized as follows.
 In Sec.~\ref{sec:theory} we minutely investigate the dynamics and the energy of the $\phi$CDM models.
 In Sec.~\ref{sec:growth} we study the influence of the $\phi$CDM models on the growth factor of matter density perturbations. In Sec.~\ref{sec:observ} presented the comparison of the obtained theoretical results with observational data.  We discuss our results and conclude in Sec.~\ref{sec:conclusion}. We use the natural units with $c= {\hbar}=1$ throughout this paper.

\section{Background dynamics in $\phi$CDM models}
\label{sec:theory}

\subsection{Background equations}
\label{sec:backeq}
We assume the presence of a self-interacting scalar field $\phi$ minimally
coupled to gravity on cosmological scales. The action of this scalar field is given by
\begin{equation}
S=\frac{M_{\rm pl}^2}{16\pi}
\int{d^{4}x\Bigl[\sqrt{-g}
\Bigl(\frac{1}{2}g^{\mu\nu}\partial_\mu\phi\partial_\nu\phi-
V(\phi)\Bigr)\Bigr]},
\label{eq:S}
\end{equation}
where $M_{\rm pl} =G^{-1/2}$ is the Planck mass with $G$ - Newtonian gravitational constant;
$V(\phi)$ is
the field's potential. Note in this presentation the scalar field $\phi$ is dimensionless, and the potential $V(\phi)$ has the $M_{\rm pl}^2$ dimension.
Following  \cite{Ratra:1987rm} we will assume that the self-interacting potential has
a power-law functional form:
\begin{equation}
V=\frac{\kappa}{2}M_{\rm pl}^2\phi^{-\alpha},
\label{eq:Potential1}
\end{equation}
where $\alpha>0$ is a model parameter, that determines the steepness of
the scalar field potential. Compliance with current observational data requires $\alpha \leq 0.7$ \cite{Samushia:2009dd}.
The larger value of $\alpha$ induces the stronger time dependence of the equation of state
parameter $ w_{\phi}$, while $\alpha$=0 corresponds to the $\Lambda$CDM case. Another model parameter
$\kappa>0$ is positive dimensionless constant which is related to $\alpha$,
(see Appendix  and Ref. \cite{Farooq:2013syn} for its dependence on $\alpha$).

We assume the flat and isotropic Universe that is described by
the standard Friedmann-Lema\^\i tre-Robertson-Walker
homogeneous cosmological spacetime model (FLRW)
$
ds^2=-dt^2+a(t)^2d{\bf x}^2$, and
we normalize the scale factor to be equal to one at present time, $a_{\rm today}=a_0=1$, i.e. $a=1/(1+z)$ where z is the redshift.

Using action for the scalar field,  Eq.~(\ref{eq:S}) we obtain
the
Klein-Gordon equation (equation of motion) for the scalar field
\begin{equation}
\ddot{\phi}+3H{\dot\phi}+\frac{\partial V(\phi)}{\partial \phi}=0,
\label{eq:KleinGordon}
\end{equation}
\noindent
where over-dot represents the derivative with the respect of
physical time $t$, $H(a)=H_0 E(a)= {\dot a}/a$ is the Hubble parameter and $H_0$ is its value today.

The flatness of the Universe requires
that the total energy density of the Universe is equal to the critical energy density, i.e. $\rho_{\rm tot}$ = $\rho_{\rm cr} $ = $3H_0^2 M_{\rm pl}^2/ (8\pi)$.
We also introduce energy density parameters for each components as, $\Omega_i = \rho_i/\rho_{\rm cr}$ (where $i$ index denotes the individual components, such as radiation, matter or the scalar field).

The energy density and pressure of the scalar field are given by
\begin{eqnarray}
\rho_\phi & = &\frac{M_{\rm pl}^2} {32\pi} \Bigl(\dot{\phi}^2/2 + V(\phi) \Bigr),
\label{eq:Rho}  \\
P_\phi & = & \frac{M_{\rm pl}^2}{32\pi} \Bigl(\dot{\phi}^2/2 - V(\phi) \Bigr),
\label{eq:P}
 \end{eqnarray}
 The corresponding equation of state is given by
$
w=({\dot\phi}^2/2 - V(\phi))/({\dot\phi}^2/2 + V(\phi)).
$
It is clear that the requirement that $w_{\rm today} \simeq -1$
imposes constraint
${\dot \phi}^2/2 \ll V(\phi)$.

The first Friedmann
equation implies:
\begin{equation}
E^2(a)=  \Omega_{\rm r0} a^{-4} + \Omega_{\rm m0} a^{-3} + \Omega_\nu (a) + \Omega_\phi (a),
\label{eq:Friedmann}
\end{equation}
where $\Omega_{\rm r0}$ and $\Omega_{\rm m0}$ are the radiation and matter (including all non-relativistic components, except neutrinos which were relativistic at the early stages) density parameters today, while $\Omega_\nu$ is the total neutrino energy density which scales as $\propto a^{-4}$ before neutrinos non-relativization, and thereafter evolves as $a^{-3}$. The scalar field energy density parameter is given by
\begin{equation}
\Omega_{\rm \phi}(a) =\frac{1}{12H_0^2}(\dot{\phi}^2+\kappa M_{\rm pl}^2\phi^{-\alpha}),
\label{eq:omegafi}
\end{equation}
To insure the flatness of the Universe, we require that $\Omega_{m0} + \Omega_{\nu 0} = 1- \Omega_{\phi 0}$, where $\Omega_{\nu 0}$ and $\Omega_{\phi 0}$ are the current energy density parameters for neutrinos and the scalar field  respectively. Since in standard cosmological scenario the neutrino
density is believed to be negligible compared to the matter and DE densities at low redshifts, we
 will ignore this component in our computations from now on
 (as well as we neglect the radiation contribution to the today energy density).

  \subsubsection{Initial conditions}
\label{sec:incond}
We integrate the set of equations Eq.~(\ref{eq:KleinGordon}) and
Eq.~(\ref{eq:Friedmann}) numerically, starting from a very early moment
$a_{\rm in}=5*10^{-5}$ to the present time $a_0=1$. For the scalar field we assume the
following initial conditions:

\begin{eqnarray}
 \phi_{\rm in} &=&\left[\frac{1}{2}\alpha(\alpha+2)\right]^{1/2}a_{\rm in}^{\frac{4}{\alpha+2}},
 \label{eq:phi0}
 \\
{ \phi}_{\rm in}^\prime &=&\Bigl(\frac{2\alpha}{\alpha+2}\Bigl)^{1/2}a_{\rm in}^{\frac{2-\alpha}{2+\alpha}},
 \label{eq:phi0div}
 \end{eqnarray}
\noindent
where a prime denotes differentiation with respect to the scale factor a. We also used $a(t) \propto t^{1/2}$ as consistent with a radiation dominated
epoch.  These initial conditions were derived from   Eq.~(\ref{eq:KleinGordon})
(for details see Appendix A). We fix the values of parameters
$\Omega_{m0}=0.315$, $\Omega_{\phi0}=0.685$,  $h=0.673$ to the best-fit values
obtained by Planck collaboration \cite{Ade:2013zuv}.

\subsubsection{The results of computations of the dynamics and the energy  of the $\phi$CDM model.}
\label{sec:result}
\begin{figure}[t]
\resizebox{0.5\textwidth}{!}{%
 \includegraphics{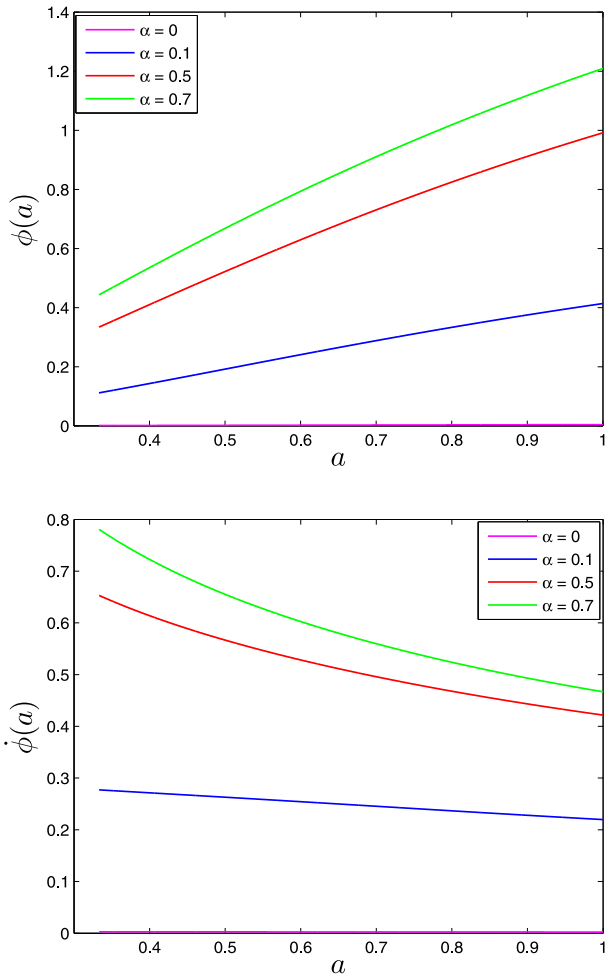}}
  \caption{The scalar field amplitude $\phi$(a) (top panel) and its time-derivative $\dot{\phi}(a)$ (bottom panel) for different values of $\alpha$ parameter.}
\label{fig:f1}       
\end{figure}

\begin{figure}[t]
\resizebox{0.5\textwidth}{!}{%
 \includegraphics{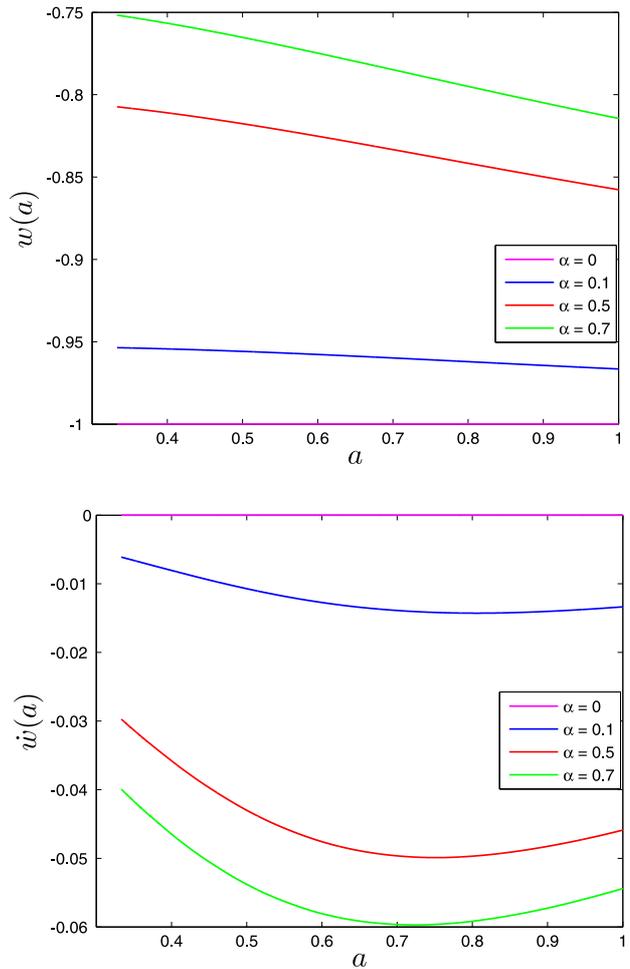}}
  \caption{DE equation of state parameter $w$(a) (top panel) and its time-derivative $\dot{w}(a)$ (bottom panel) as a function
    of scale factor for different values of $\alpha$ parameter.}
 \label{fig:f2}
\end{figure}

\begin{figure}[t]
\resizebox{0.5\textwidth}{!}{%
 \includegraphics{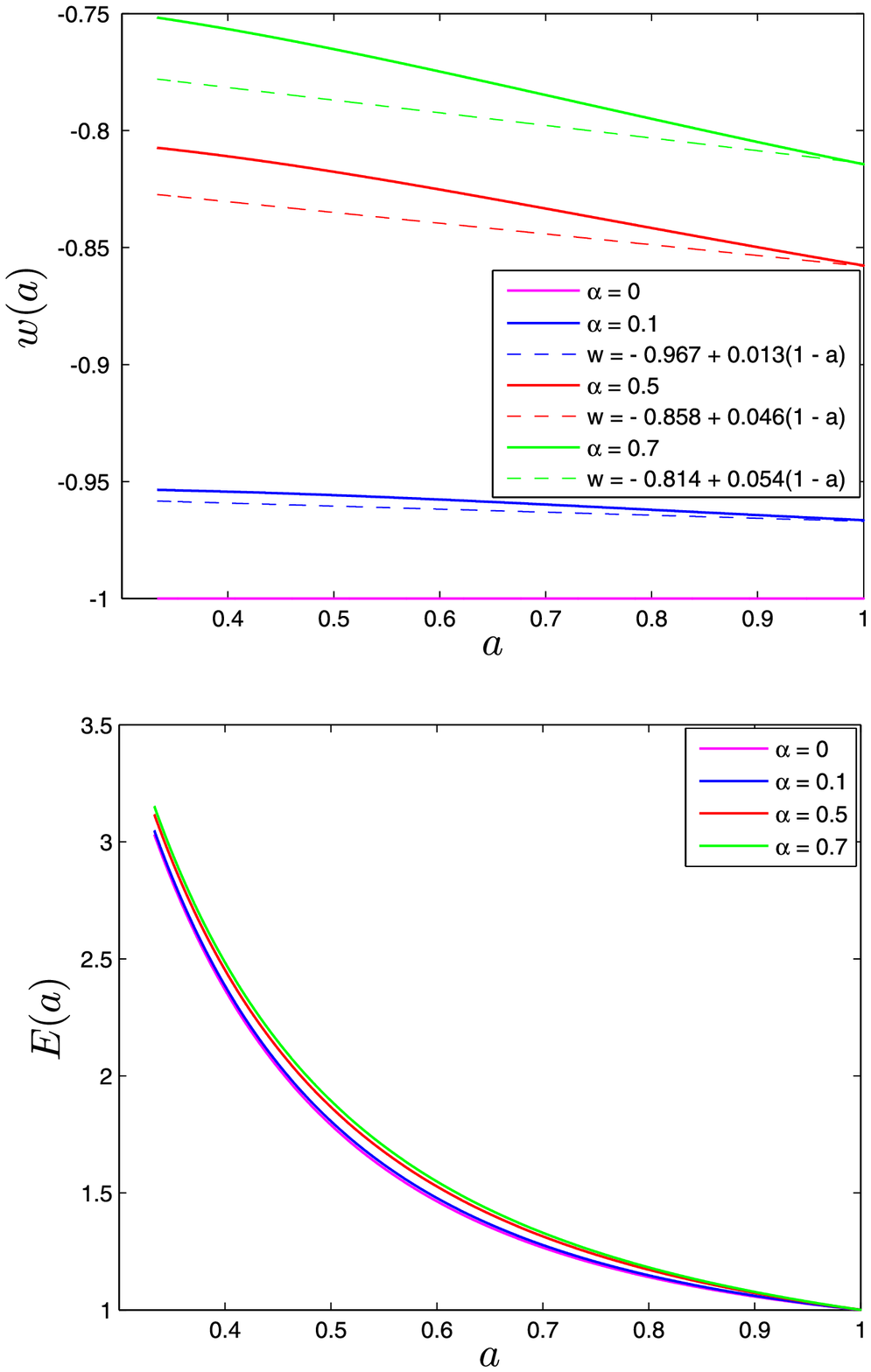}}
  \caption{On the top panel is shown $w(a)$ for different values of $\alpha$ parameter along with predictions computed from the CLP parametrization with corresponding
    best-fit values for $w_0$ and $w_\mathrm{a}$. On the bottom panel is shown the normalised Hubble expansion rate $E(a)$ for different model parameter $\alpha$.}
 \label{fig:f3}
\end{figure}

\begin{figure}[t]
\resizebox{0.5\textwidth}{!}{%
 \includegraphics{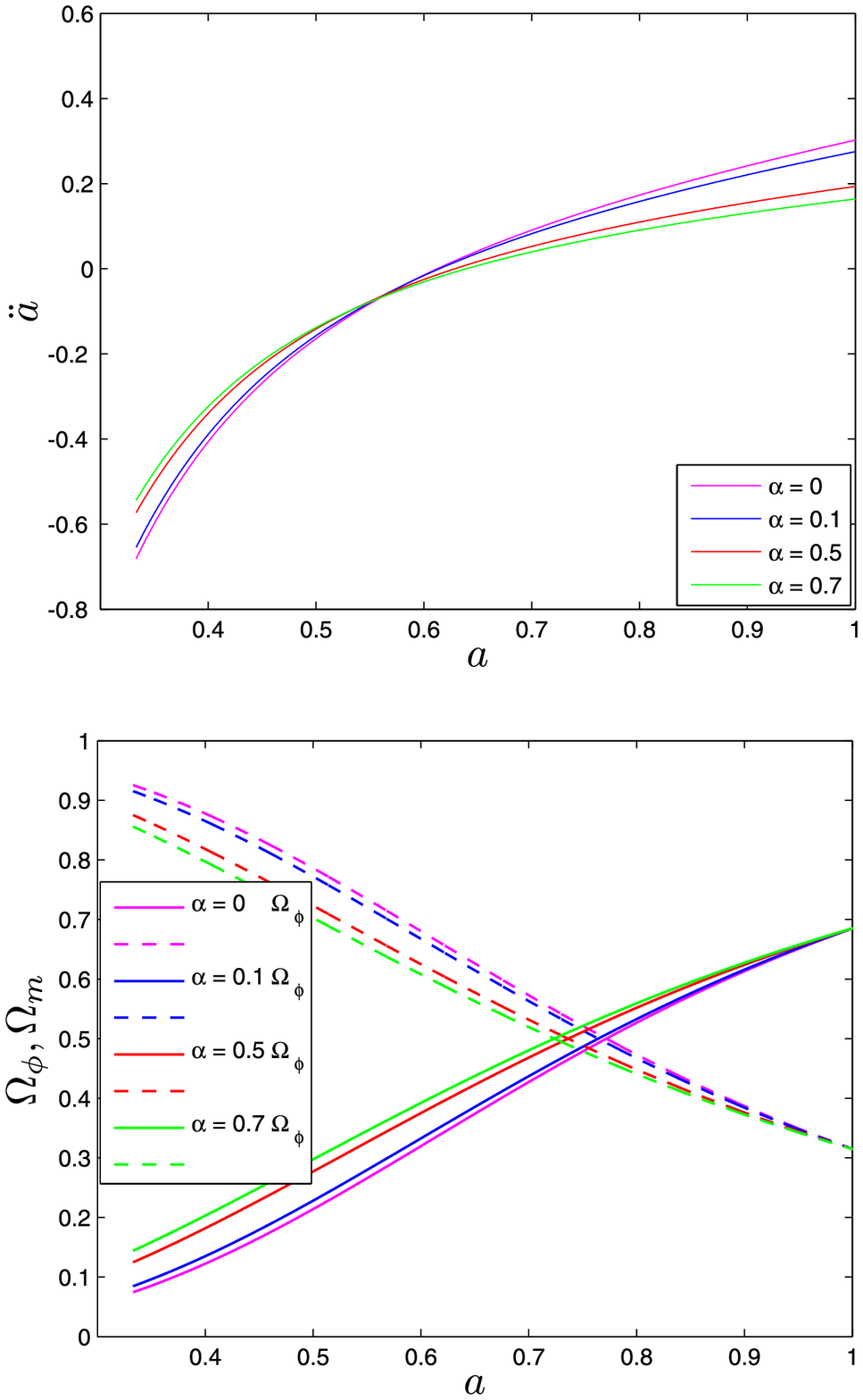}}
 \caption{ The second derivative of the scale
factor (top panel) and energy densities of $\Omega_m$(a) (dashed lines) matter and $\Omega_{\rm \phi}$(a) (solid lines) scalar field (bottom panel) as functions
of scale factor for different values of $\alpha$ parameter.}
 \label{fig:f4}
\end{figure}

We present the background
dynamics in the presence of scalar field DE on Figs.~(\ref{fig:f1})-(\ref{fig:f4}).
The large values of the $\alpha$ parameter imply larger values of the scalar field amplitude $\phi(t)$ and its time  derivative $\dot{\phi}(t)$  at
all redshifts. The large values of  the $\alpha$ parameter result also in the large values of $w$ and
$dw/da$ at all redshifts.

The evolution of the equation of state $w(a)$ is presented on Fig.~\ref{fig:f3}. We find that for all values of
the $\alpha$ parameter, the Chevallier-Polarsky-Linder (CPL) parametrization of DE
equation of state  $w(a)=w_0+w_a(1-a)$  near $a=1$ (where  $w_0=w(a=1)$ and $w_a=(-dw/da)|_{a=1}$) \cite{Chevallier:2000qy} provides a good approximation
in the range of scale factor  $a = [0.98-1]$.

The evolution of $E(a)$ for different values $\alpha$ parameters is shown on
Fig.~\ref{fig:f3}. As we can expect the expansion of the Universe occurs more rapidly with
increasing value of the $\alpha$  parameter, the $\Lambda$CDM limit corresponding
to the slowest rate of the expansion. The value of the
$\alpha$ parameters affects also the redshift of the equality between matter and scalar
field energy densities (see Fig.~\ref{fig:f4}.1); with larger values of the $\alpha$ the scalar field
domination begins earlier and vice versa.

\section{Growth factor of matter density perturbations in dark energy models}
\label{sec:growth}

We use the linear perturbation equations
for matter  overdensities \cite{Pace:2010sn,Campanelli:2011qd}
to describe the evolution of small overdensities in a homogeneous expanding
Universe,
\begin{equation}
\delta^{''}+\Bigl(\frac{3}{a}+\frac{E^{'}}{E}\Bigr)\delta^{'}-\frac{3\Omega_{m,0}}{2a^{5}E^{2}}\delta=0,
\label{eq:deltaeq}
\end{equation}
where $\delta \equiv \delta\rho_m/\rho_m$,
with $\rho_m$ and $\delta\rho_m$ the
density and overdensity of the matter component, respectively.
Following \cite{Pace:2010sn} we use the initial conditions $\delta(a_{\rm in})=\delta^{'}(a_{\rm in})=5*10^{-5}$,
with $a_{\rm in}=5*10^{-5}$ as defined above.

We define
$
D(a)=\frac{\delta(a)}{\delta(a_i)}
$
 - the linear growth rate, that shows how much the
perturbations have grown since initial moment $a_{\rm in}$.  We normalize the growth
rate, so that $D(a_{\rm in})=1$.
The fractional matter density $f_1(a) \equiv \Omega_m(a) $ as a function of time is given by
$
f_1(a)=\Omega_{\rm m0}a^{-3}/E^2,
$
and we define the function $f_2(a)$, that describes the growth rate of the matter perturbations
as a logarithmic derivative of linear growth rate \cite{Wang:1998gt}:
$
f_2(a) ={dlnD(a)}/{dlna},
$
in $\Lambda$CDM cosmology the two functions can be related as
\begin{equation}
 f_2(a) \approx [f_1(a)]^{\gamma},
 \label{eq:f1f2}
\end{equation}
The $\gamma$ parameter is also referred to as a growth index
\cite{Linder:2005in}, and it depends on both a model of DE and a theory of
gravity. In general relativity (GR) the time dependence of the $\gamma$ index can
be fitted by \cite{Linder:2005in}:
\begin{equation}
\gamma=0.55+0.05(1+w_0+0.5w_a),
~~{\rm if}~~ w_0 \ge-1.
\label{eq:gamma}
\end{equation}
\noindent
For $\Lambda$CDM model (with $w=-1$), the growth index $\gamma = 0.55$ \cite{Linder:2005in,Samushia:2010ki}.
$\phi$CDM model has been tested through the growth rate in Ref. \cite{Pavlov:2013uqa}. In more complex coupled dark energy models, the growth rate has been studied in Refs. \cite{Neupane:2009wc,Piloyan:2014gta,Song:2008qt}. The measured value of $\gamma$ in conjunction with tight constraints on other cosmological parameters, can be used to test the validity of GR, see Refs. \cite{Taddei:2014wqa,Pouri:2014nta} for recent studies to use  the linear growth rate data to
determine the deviation of the theory of gravity on extragalactic
scales from the standard GR.

\subsection{The results of  computations of the growth factor of matter density perturbations in $\phi$CDM dark energy model}

\begin{figure}[t]
\resizebox{0.5\textwidth}{!}{%
 \includegraphics{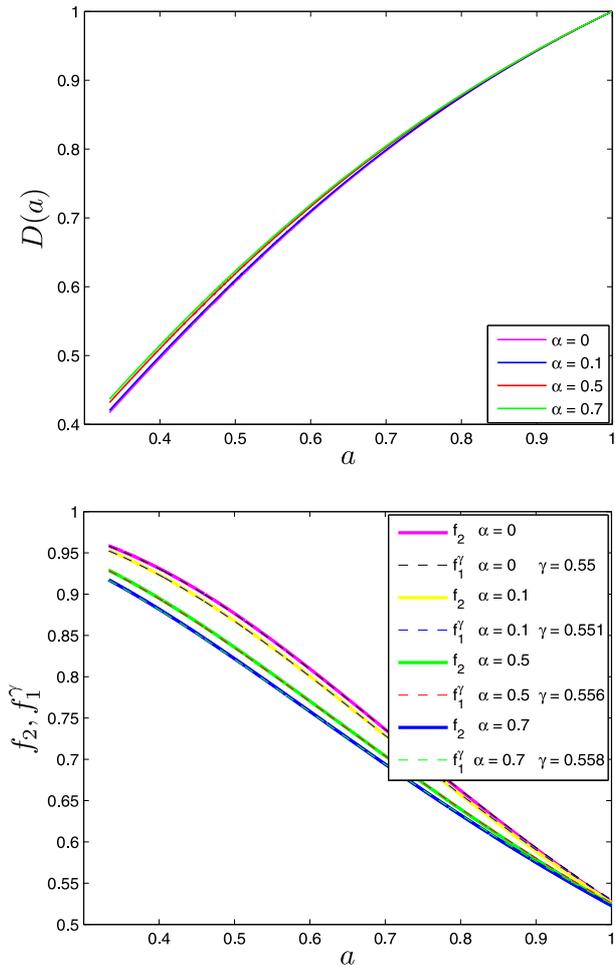}}
\caption{ On the top panel is shown the linear growth  as D(a) as a function of scale factor for different values of $\alpha$ parameter. On the bottom panel is shown the logarithmic growth rate as a function  of scale factor for different values of $\alpha$ parameter $f_2$ (solid lines)  along with the predictions $f_1^\gamma$ (dashed lines), computed for the corresponding best-fit values of $\gamma$ parameter.} 
\label{fig:f5}
\end{figure}

We present the solutions of the growth equation Eqs.~(\ref{eq:deltaeq})
in RP models on Fig.~(\ref{fig:f5}).

We have checked that the power-low approximation Eqs.~(\ref{eq:f1f2})
works well for the
scalar field DE. The effective value of the growth index $\gamma$ depends on $\alpha$
and is slightly higher than the $\Lambda$CDM limit of 0.55.

\begin{figure}[t]
\resizebox{0.5\textwidth}{!}{%
 \includegraphics{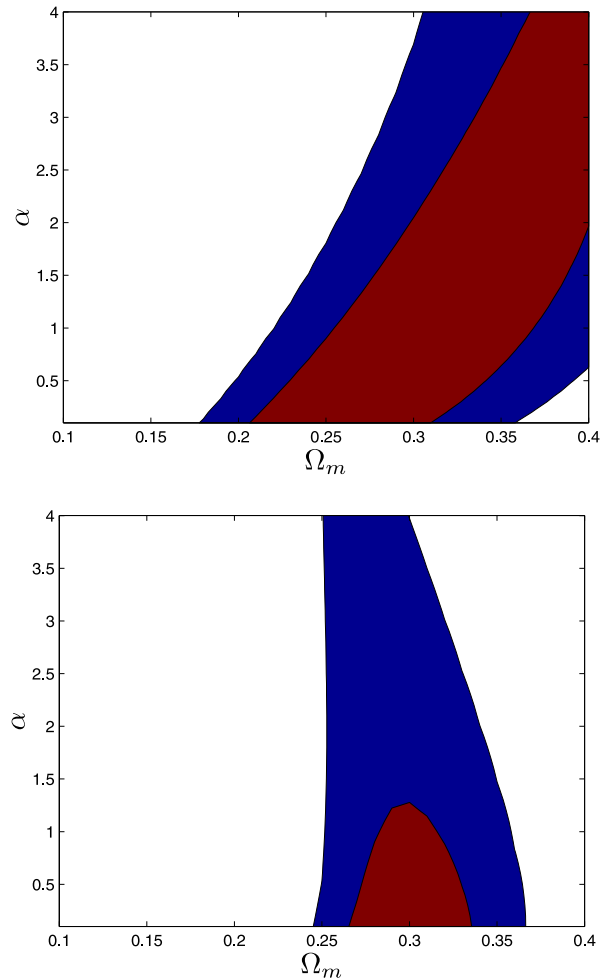}}
 \caption{ 1$\sigma$  and  2$\sigma$  confidence level contours on parameters
$\Omega_{\rm m}$ and $\alpha$ of $\phi$CDM model. On the top panel is shown
constraints, obtain from the growth rate data \cite{Gupta:2011kw}.
On the bottom panel is shown constraints, obtained after adding BAO
measurements and CMB distance prior as in \cite{Giostri:2012ek} for
BAO/CMB distance prior.}
 \label{fig:f6}
\end{figure}

\section{Comparison with observations}
\label{sec:observ}

The
$\phi$CDM models generically predict a faster expansion rate and a slower rate of
growth at low redshifts. Tight measurements of the expansion rate,
distance-redshift relationship and the growth rate at multiple redshift ranges
can be used to simultaneously constrain the background dynamics and the growth
of structure and discriminate between $\phi$CDM and $\Lambda$CDM models.

For the rest of this section we will concentrate specifically on the
discriminative power of the growth rate and BAO measurements from galaxy
surveys. For simplicity we will assume that the spatial-curvature is known
precisely and $\Omega_{\rm k} = 0$.  \cite{Pavlov:2013nra} explored in detail
the background dynamics and the growth of structure of the generalize non-flat
$\phi$CDM model. We take  a compilation of growth rate measurements from
\cite{Gupta:2011kw} and obtain posterior likelihood function of
parameters $\alpha$ and $\Omega_\mathrm{m}$. To do this we apply the same method as \cite{Gupta:2011kw};
we numerically solve Eq.~(\ref{eq:deltaeq}) for series of $\alpha$ and $\Omega_\mathrm{m}$ values and
compute a $\chi^2$ value
\begin{equation}
\chi^2(\alpha,\Omega_\mathrm{m}) = \frac{[f_\mathrm{m} - f_\mathrm{th}(\alpha,\Omega_\mathrm{m})]^2}{\sigma_f^2},
\end{equation}
\noindent
where $f_\mathrm{m}$ is a measured value of growth rate, $f_\mathrm{th}$ a theoretically computed value
and $\sigma_f^2$ one standard deviation error of the measurement.
Assuming that the likelihood is Gaussian we have
\begin{equation}
\mathcal{L}^\mathrm{f}(\alpha,\Omega_\mathrm{m})\propto \mathrm{exp}[-\chi^2(\alpha,\Omega_\mathrm{m})/2].
\end{equation}
The 1 and 2$\sigma$ confidence
contours resulting from this likelihood are presented on the top panel of
Fig.~\ref{fig:f5}. The likelihood contours in $\alpha$ - $\Omega_{\rm m}$ plane
obtained from the growth rate data alone are highly degenerate. If we fix
$\alpha=0$ we get $\Omega_\mathrm{m} = 0.278 \pm 0.03$ which is within
1$\sigma$ of the best-fit value obtained by Planck collaboration
\cite{Ade:2013zuv}. Values of $\Omega_{\rm m} < 0.2$ are ruled out at more
than $2\sigma$ confidence level, but large values of $\Omega_{\rm m}$ are still
allowed as long as the $\alpha$ is large.

To break the degeneracy between $\Omega_\mathrm{m}$ and $\alpha$ parameters we
now add a compilation of low-redshift BAO measurements from
\cite{Giostri:2012ek}. We follow the same approach as \cite{Giostri:2012ek};
we compute angular distance
\begin{equation}
d_\mathrm{A}(z, \alpha, \Omega_\mathrm{m}, H_0) = c\int_0^z \frac{dz'}{H(z', \alpha, \Omega_\mathrm{m}, H_0)},
\end{equation}
\noindent
and a distance scale

\begin{multline}
D_\mathrm{V}(z, \alpha, \Omega_\mathrm{m}, H_0) = \\
 [d_\mathrm{A}^2(z, \alpha, \Omega_\mathrm{m}, H_0)cz/H(z, \alpha, \Omega_\mathrm{m}, H_0)]^{1/3},
\end{multline}
\noindent
at a series of redshifts and construct a combination $\eta(z) \equiv d_\mathrm{A}(z_\mathrm{bao})/D_\mathrm{V}(z_\mathrm{bao})$
where $H(z)$ is the Hubble parameter and $H_0$ is a Hubble constant. Assuming Gaussianity of the errorbars
we again compute the $\chi^2$
\begin{equation}
\chi^2_\mathrm{bao} = \bm{X}^\mathrm{T}\bm{C}^{-1}\bm{X}
\end{equation}
\noindent
and a likelihood function
\begin{equation}
\mathcal{L}^\mathrm{bao}(\alpha,\Omega_\mathrm{m},H_0) \propto \mathrm{exp}(-\chi^2_\mathrm{bao}/2),
\end{equation}
\noindent
where $\bm{X} = \eta_\mathrm{th} - \eta_\mathrm{m}$ and $\bm{C}$ is the
covariance matrix of the measurements.  To marginalize over parameter $H_0$ in
$\mathcal{L}^\mathrm{bao}$ we take a Gaussian prior of $H_0 = 74.3 \pm 2.1$
from \cite{Freedman:2012}. We assume that $\mathcal{L}^\mathrm{f}$ and
$\mathcal{L}^\mathrm{bao}$ are independent and the combined likelihood is
simply a product of the two. The results are presented on the bottom panel of
Fig.~\ref{fig:f5}. The addition of BAO measurements breaks degeneracy in the
growth rate data.  $\Omega_{\rm m}$ is now constrained to be within $0.26 <
\Omega_{\rm m} < 0.34$ at 1$\sigma$ confidence level. For $\alpha$ parameter we
get $0\leq\alpha\leq1.3$ at 1$\sigma$ confidence level.

\section{Discussion and Conclusions}
\label{sec:conclusion}

We explored observable predictions of scalar field DE model.  We
showed that the model differs from $\Lambda$CDM in number of ways that are
generic and do not depend on the specific values of model parameters.  For
example, in scalar field models the expansion rate of the Universe is always
faster and the DE dominated epoch sets in earlier than in $\Lambda$CDM
model when other cosmological parameters are kept fixed.
The two models also differ in their predictions for the growth rate, where the
scalar field model generically predicts a slower growth rate than $\Lambda$CDM.

We used a compilation of BAO, growth rate and the distance prior from the CMB
to constrain model parameters of the scalar field model. We find that if only
the growth rate data is used there is a strong degeneracy between
$\Omega_\mathrm{m}$ and $\alpha$, where higher values of $\alpha$ are allowed
as long as the $\Omega_\mathrm{m}$ parameter is large as well. When combining
these constraints with the constraints coming from a distance-redshift
relationship (BAO data and the distance prior from CMB) the degeneracy is
broken and we get $\Omega_\mathrm{m} = 0.30 \pm 0.04$ and $\alpha < 1.30$ with
a best fit value of $\alpha = 0.00$.

{\it Acknowledgements}
We appreciate useful comments from Leonardo Campanelli. We thank Gennady Chitov, Omer Farooq, Vasil Kukhianidze, Anatoly Pavlov, Bharat Ratra,  and Alexander Tevzadze for discussions.  We
acknowledge partial support from the Swiss NSF grant SCOPES IZ7370-152581, the
CMU Berkman foundation, the NSF grants AST-1109180, and the
NASA Astrophysics Theory Program grant NNXlOAC85G. N. A. and T.K. acknowledge hospitality of International Center for Theoretical Physics (ICTP, Italy) where this work has been designed.

\vskip20pt
\appendix{\bf{APPENDIX A: Calculation of $\kappa$ factor\\}}
\noindent
In the appendix we calculate the $\kappa$ factor following Sec.~3.6.3 of Ref. \cite{Farooq:2013syn}.
Let's represent the scale factor and the scalar field $\phi(t)$ in the power-law forms,
\begin{equation}
a(t)=a_\star \Bigl(\frac{t}{t_\star}\Bigl)^n,~~~~~~~\phi(t)= \phi_\star \Bigl(\frac{t}{t_\star}\Bigl)^p
\label{eq:Power}
\end{equation}
where $a_\star \equiv a(t_\star)$ and $\phi_\star \equiv \phi(t_\star)$ are the scale factor and the scalar field values at $t=t_\star$.
Eq.~(\ref{eq:KleinGordon}) implies $p=2/(2+\alpha)$ (see for details Sec.~3.6.3 of Ref. \cite{Farooq:2013syn}), and respectively,
\begin{equation}
\phi_\star^{\alpha+2}=\frac{(\alpha+2)^2}{4(6n+3n\alpha-\alpha)}\kappa
\alpha M_{\rm pl}^2 t_\star^2.
 \label{eq:Scalmod2}
\end{equation}
Using Eq.~(\ref{eq:Power}) with Eq.~(\ref{eq:Scalmod2}) with Eq.~(\ref{eq:Rho}) and Eq.~(\ref{eq:Friedmann}), we obtain
\begin{eqnarray}
\rho & = &   \frac{3n}{8\pi}\Bigl(\frac{M_{\rm pl}}{t_\star} \Bigl)^2  \frac{\phi_\star^2}{\alpha(\alpha+2)}
\Bigl(\frac{t}{t_\star}\Bigl)^{\frac{-2\alpha}{\alpha+2}}
 \label{eq:Rho1}
\\
\Bigl(\frac{n}{t}\Bigl)^2  &= &\frac{8\pi}{3M_{\rm pl}^2}\rho,
\label{eq:Rho2}
\end{eqnarray}
where $\rho \equiv \rho_\phi$ and we assume that it is the energy density of a single component that
was dominant at $t<t_\star$ in the Universe. Assuming
 $
 \rho(t) = \rho_\star ({t}/{t_\star})^\beta
 $
 we get
 $\beta = -2\alpha/(\alpha+2)$. On the other hand,
 assuming that the component energy density is $\rho_\star$ at $a=a_\star$, and accounting for the dominance of the component at this epoch,
we have
\begin{equation}
\rho=\rho_{\star}\Bigl(\frac{a_\star}{a}\Bigl)^{\frac{2}{n}},
 \label{eq:Rho3}
\end{equation}
where $n=1/2$ is for radiation and $n=2/3$ for the matter dominant epochs.
 Expressing $1/t^2$ through Eq.~(\ref{eq:Rho2}),  and using Eq. (\ref{eq:Rho3}) in Eq.~(\ref{eq:Rho1}) with assuming $a=a_\star$, $\rho=\rho_{\star}$, we can derive $\phi_\star^2$
and comparing
the obtained result with Eq.~(\ref{eq:Scalmod2}), we find
\begin{equation}
\kappa=\frac{32 \pi}{3nM_{\rm pl}^4}
\Bigl(\frac{6n+3n\alpha-\alpha}{\alpha+2}\Bigl)[n\alpha(\alpha+2)]^{\frac{\alpha}{2}}
\rho_{\star }.
 \label{eq:kappa}
\end{equation}
Plugging Eq.~(\ref{eq:kappa}) in Eq.~(\ref{eq:Scalmod2}), and using  Eq.~(\ref{eq:Rho2}),
we obtain
\begin{eqnarray}
\phi_\star &= &[n\alpha (\alpha+2)]^{\frac{1}{2}},
\label{phi2}
\\
\phi&=&[n\alpha (\alpha+2)]^{\frac{1}{2}} \Bigl(\frac{a}{a_\star}\Bigl)^{\frac{2}{n(\alpha+2)}}
\label{phi3}
\end{eqnarray}
that simultaneously lead to initial conditions, Eq.~(\ref{eq:phi0}) and Eq.~(\ref{eq:phi0div}) in the radiation dominated epoch with $n=1/2$ through assuming $a_\star=a_0$.

Plugging Eq.~(\ref{phi3}) in Eq.~(\ref{eq:KleinGordon}),
\begin{equation}
\kappa=\frac{4n}{M_{\rm pl}^2 t_\star^2}
\Bigl(\frac{6n+3n\alpha-\alpha}{\alpha+2}\Bigl)[n\alpha(\alpha+2)]^{\alpha/2},
 \label{eq:kappa1}
\end{equation}
Since Eq.~(\ref{eq:kappa}) must be valid for any $t_\star$, we imply the freedom of our choice and use $t_\star=M_{\rm pl}^{-1}$. Finally
we have for $n=1/2$ and $n=2/3$ respectively:
\begin{eqnarray}
\kappa(n=1/2) & = & \Bigl(\frac{\alpha+6}{\alpha+2}\Bigl)
\Bigl[\frac{1}{2}\alpha(\alpha+2)\Bigl]^{\alpha/2},
\label{eq:kappa2}
\\
\kappa(n=2/3) & = & \frac{8}{3} \Bigl(\frac{\alpha+4}{\alpha+2}\Bigl)
\Bigl[\frac{2}{3}\alpha(\alpha+2)\Bigl]^{\alpha/2}.
 \label{eq:kappa3}
\end{eqnarray}

\end{document}